# Scaling Theory of Two-Dimensional Field Effect Transistors

Saurabh V. Suryavanshi, Chris D. English, H.-S. P. Wong, and Eric Pop

Department of Electrical Engineering, Stanford University
Stanford, CA 94305, USA

**Abstract:**

We present a scaling theory of two-dimensional (2D) field effect transistors (FETs). For devices with channel thickness less than 4 nm, the device electrostatics is dominated by the physical gate oxide thickness and not the effective oxide thickness. Specifically, for symmetric double gate (DG) FETs the scale length ($\Lambda$) varies linearly with the gate oxide thickness ($t_{ox}$) as $\Lambda \sim \frac{3}{4} t_{ox}$. The gate oxide dielectric permittivity and the semiconductor channel thickness do not affect the device electrostatics for such device geometries. For an asymmetric device such as single gate (SG) FETs, the fringing fields have a second order effect on the scale length. However, like symmetric DG FETs, the scale length in asymmetric FETs is also ultimately limited by the physical gate oxide thickness. We compare our theoretical predictions for scaled monolayer $MoS_2$ DG FETs.

# I. INTRODUCTION

Moore's law [1] has encouraged the steady reduction in the size of field effect transistors (FET) for decades, and this has followed Dennard's recommendation [2] at least until 2005. After 2005, these "geometric scaling" trends have slowed down, but progress was sustained by "equivalent scaling" innovations [3], such as the introduction of metal gates with high-κ dielectrics, FinFET (or tri-gate) geometries, and strain engineering.

In extremely small FETs, the drain contact is physically closer to the source contact and influences the injection of carriers into the channel, which is undesirable. In other words, FETs with short channel length ($L$) are more difficult to turn off using the gate voltage, and the subthreshold leakage currents are increased. The minimum channel length for a given transistor technology is limited to a multiple of a scale length $\Lambda$ [4][5], for example ~3$\Lambda$ for planar bulk FETs [4]. Transistors with shorter channels would suffer from so-called short channel effects (SCEs), which include drain induced barrier lowering (DIBL), increased subthreshold slope (*SS*), and threshold voltage ($V_T$) roll-off.

Extensive research into FET electrostatics has pointed out that $\Lambda$ depends on the channel thickness ($t_{ch}$), and specifically that a smaller $t_{ch}$ yields a shorter $\Lambda$. This has driven significant efforts to replace bulk Si, Ge, or III-V FETs with ultrathin semiconductor-on-insulator (SOI) transistors [6]–[8], FinFETs or tri-gate transistors [9][10], and surround-gate transistors [11], including recent multi-stacked nanosheet designs [12]. However, in all such devices based on bulk semiconductors, the channel thickness is greater than the gate oxide thickness, $t_{ch} > t_{ox}$.

In contrast to bulk semiconductors, the atomic limit of transistor scaling could be enabled by semiconducting two-dimensional (2D) materials and devices [13]. These materials are intrinsically made of single molecular layers with a thickness less than 1 nm [14]. For example, one of the most studied 2D materials, $MoS_2$, has a single layer thickness of three atoms or 6.15 Å. These 2D FETs can be viewed as an extreme example of SOI technology with an atomically thin channel and the special case of $t_{ch} < t_{ox}$. However, unlike ultra-thin SOI these 2D materials retain excellent transport properties even at small channel thickness [10], [15], yet their device electrostatics and specifically their scale length have not been studied in depth.

In this work, we employ finite element simulations to analyze the device electrostatics for 2D FETs with an extremely thin channel. We show that for symmetric double-gate (DG) FETs with $t_{ch} < 4$ nm, the scale length is mostly independent of $t_{ch}$ and gate oxide dielectric constant ($\varepsilon_{ox}$). In this limit of an extremely thin channel, the dominant determining factor for the scale length is the physical thickness of the gate oxide ($t_{ox}$). This sharply contrasts with previous findings regarding bulk semiconductor FETs (including SOI) that $\varepsilon_{ox}$ and $t_{ch}$ are the primary variables for reducing

$\Lambda$. Instead, we show that the gate oxide dielectric constant plays a secondary and not a dominant role in determining SCE such as DIBL. We also analyze single gate (SG) FETs based on 2D semiconductors and elucidate the effect of drain fringing fields (also known as fringing field induced barrier lowering or FIBL [16]) on the scale length. Furthermore, we compare our theory with experimental data for monolayer MoS$_2$ DG FETs and show that our simulations agree with the existing experimental results. Using the experiments and simulations, we also highlight the importance of the contact architecture on device electrostatics.

## II. Methodology

The sample device geometries are shown in Figs. 1a and 1b. The source and drain contacts effectively act as edge contacts to the 2D channel and extend over the thickness of the gate oxides. We refer to them as tall contacts and the height of the tall contact to DG FET is $t_c = 2t_{ox} + t_{ch}$. Note that such tall contacts capture the worst-case impact as the drain field can penetrate the device through a larger cross-section area. We later also discuss short contacts in which the contact is only made to the semiconductor channel and not the side of the gate oxide, representing real devices. For short contacts, $t_c = t_{ch}$. The contacts are assumed to be Ohmic in all simulations.

We are primarily interested in device electrostatic simulation in the absence of charge carriers in the channel. This involves solving Poisson's equations inside the device using commercially available tools [17]. As explained later, we do not simulate carrier transport and therefore do not need to include quantum effects. We capture the SCE by simulating devices in the sub-threshold region of operation ($V_{GS} < V_T$). For these undoped devices ($V_T \approx 0$ V), $V_{GS}$ = -1 V is maintained in all simulations to keep the device in the sub-threshold regime. Channel lengths are incremented from $L$ = 5 nm to 500 nm to observe the onset of the SCE. For each configuration (with an $L$, $t_{ch}$, $t_{ox}$, and other device parameters), we extract the minimum electrostatic potential ($\psi_{s,min}$) along the channel (x-direction, white dashed lines, Figures 1a-b).

Figure 1c shows electrostatic potential $\psi_s$ and $\psi_{s,min}$ for a DG FET with $L$ = 40 nm, $t_{ox}$ = 4 nm, and $t_{ch}$ = 0.6 nm. (Thickness of a 1L of MoS$_2$ is ~ 0.6 nm.) For a symmetric DG FET, the $\psi_{s,min}$ is located exactly in the middle of the channel ($y = 0$). For the SG and asymmetric DG FETs (with different oxides on either side of the channel), $\psi_{s,min}$ needs to be extracted off-center along the path of the maximum current flow. For the long channel case ($L \gg \Lambda$), the drain is farther away from the source and $\psi_{s,min}$ is approximately constant irrespective of $L$. However, for small $L$, the onset of SCE causes $\psi_{s,min}$ to increase. In Fig. 1d, we plot the deviation of the $\psi_{s,min}$ ($\Delta\psi_{s,min}$) from that of the long-channel ($L$ = 500 nm). Note that this $\Delta\psi_{s,min}$ is representative of DIBL and increases exponentially with reduced channel length [4]. We, therefore, fit an exponential of the form ~exp(-$\Lambda/L$) to the $\Delta\psi_{s,min}(L)$ curve and extract the scale length, $\Lambda$, as shown in Fig. 1d. The extracted Λ is a constant for a particular gate stack and channel thickness and does not depend on the channel length. THE SCE such as DIBL on the other hand depends on Λ as well as the channel length.

### III. Results and Discussion

**Scale length in symmetric DG FET**. In Fig. 2, we compare the scale length obtained from simulations to previous scale length models [6]–[8]. In 2D FETs, a majority of the drain influence over the channel barrier is due to fringing fields through the gate oxide or FIBL. The effect of this fringing field has not been dominant in the literature as traditional transistor devices have had channels thicker than the gate oxide. Especially for smaller $t_{ch}$ (< 2 nm), it can be observed in Fig. 2 that Yan et al. [8] and Suzuki et al. [6] severely underestimate the scale length. These scale length studies approximated as $\Lambda \approx \sqrt{\epsilon_{ch} t_{ch} t_{ox}/\epsilon_{ox}}$ also predict a much smaller Λ as $t_{ch} \to 0$. On the other hand, our finite-element simulations show a much higher and a non-zero Λ as $t_{ch} \to 0$. Frank et al. [7] iteratively solves for electrostatic in a DG FET with tall source and drain contacts and get results similar to our simulations.

We first study the scale length in detail for the symmetric DG FET as shown in Fig. 3. The Λ is simulated for different oxide dielectric constants ($\varepsilon_{ox}$), oxide thickness ($t_{ox}$), and channel thickness ($t_{ch}$). We have kept the dielectric constant of the semiconductor ($\varepsilon_{ch}$) constant with channel thickness in all cases. In Fig. 3a, we plot Λ for a fixed oxide thickness ($t_{ox}$ = 5 nm) versus channel thickness for different oxide dielectric constants. We observe that for large $t_{ch}$, Λ reduces with increasing $\varepsilon_{ox}$. However, as we reduce $t_{ch}$, Λ converges to a single value of ~6.6 nm irrespective of the oxide dielectric constant. This shows that in the limit of $t_{ch} \to 0$, Λ is independent of $\varepsilon_{ox}$. In Fig. 3b, we plot Λ vs. $t_{ch}$ for different $t_{ox}$ while maintaining $\varepsilon_{ox}$ = 12. In this figure, we observe two distinct regimes. For thicker channels, Λ increases with $t_{ch}$. On the other hand, for thinner channels Λ is mostly independent of $t_{ch}$. Thus, for applications such as 2D FETs where $t_{ch}$ < 1nm, Λ is mostly independent of $t_{ch}$ and $\varepsilon_{ox}$. In other words, Λ depends only on the geometry of the gate oxide ($t_{ox}$). This observation is similar to electrically doped 2D Tunnel FETs [18], where the channel thickness has minimal influence on the device electrostatics.

In Fig. 3c, we plot Λ versus $t_{ox}$ for the symmetric DG FET with small $t_{ch}$ (= 0.5 nm) and $\varepsilon_{ox}$ = 12. We consider two types of contacts: the short contact (realistic case) and the tall contact (worst case). In the case of the short contact, the source and the drain terminal contact only the semiconductor channel. For the tall contact, the source and drain extend all the way to the gate oxide as shown in the Fig. 1a-b. We observe that for both types of contacts Λ is proportional to $t_{ox}$. The scale length for short contacts and tall contacts start to deviate from each other and from the linear approximation when $t_{ox}/t_{ch} > 10$. The slope, $\gamma$, represents the strength of the gate control; smaller $\gamma$ implies better gate control (small $\Lambda$) while large $\gamma$ implies poor gate control (large $\Lambda$). For 2D DG FETs, we see that $\gamma \sim 4/3$ irrespective of the gate-oxide dielectric constant.

For an ideal transistor switch, we expect the gate field to completely control the switching operation. The undesirable short channel effects such as DIBL are a consequence of competition between the drain electric field and the gate electric field to influence the channel barrier. This influence can also be thought of as an effective capacitive coupling from the drain and the gate to the channel barrier. Larger the capacitive coupling, larger is the influence of the drain or the gate. In devices with a thicker channel, the drain field influences the channel barrier through the gate oxide (also known as field induced barrier lowering or FIBL) as well as the semiconducting channel. In the case of thinner channels, the drain field affects the channel barrier predominantly through the gate oxide (FIBL). Therefore, for devices with extremely thin channels such as 2D FETs, the relative capacitive coupling between the drain and the gate to the channel barrier (near the source terminal) does not change with the gate-oxide dielectric constant. The geometry of the gate oxide (i.e. $t_{ox}$, $L$), and not the dielectric constant, controls short channel effects.

To further understand device electrostatics, we show the electrostatic potential in representative structures in Fig. 4 with $L = 10$ nm. The structure in Fig. 4a-c have $\varepsilon_{ox} = 1$ while Fig. 4d-f have $\varepsilon_{ox} = 10$. For all structures, $t_{ox}$ is fixed at 5 nm. For thinner channels (Fig. 4a and 4d), the impact of dielectric constant on the device electrostatic is negligible. For thicker channels (Figs. 4c and 4f), however, larger oxide dielectric constant reduces the relative penetration of the drain field into the channel and improves the gate control.

**Scale length in asymmetric FETs or SG FETs.** We simulate the SG FET as an extreme example of asymmetric FETs. As shown in Fig. 1b, the SG FET has a gate electrode on one side. The other side of the semiconducting channel can be air (in the case of most experimental back-gate FETs) or a dielectric with a dielectric constant of $\varepsilon_{sd}$ (such as oxide in SOI structures). In the case of an SG FET, the drain field influences the channel potential barrier through the surrounding dielectric in addition to the gate oxide.

The scale length in the asymmetric FETs such as SG FET depends on the oxide dielectric constant as well as the surrounding dielectric constant. For symmetric DG FET, the gate coupling strength $\gamma \approx 4/3$. In Fig. 5a, for a fixed $\varepsilon_{sd}$, $\gamma$ reduces with $\varepsilon_{ox}$ indicating better gate control. On the other hand, for a fixed $\varepsilon_{ox}$, the gate control degrades with increasing $\varepsilon_{sd}$ as shown in Fig. 5b. The FIBL from the drain through the surrounding dielectric increases and $\gamma$ increases. Notably, in both Figs. 5a and 5b, the device electrostatics is ultimately limited by the limit of the symmetric DG FET ($\gamma \approx 4/3$), which depends only on the gate oxide thickness.

**Discussion on DIBL and relevance of the gate oxide dielectric constant**. We showed that $\Lambda$ does not depend on $\varepsilon_{ox}$. Though DIBL has an exponential dependence on $\Lambda$ as $\propto \exp(-\Lambda/L)$, it does have a negligible dependence on gate oxide dielectric constant. In Fig. 6 we plot DIBL for various channel lengths and gate oxide dielectric constants in a DG FET. As expected, the DIBL varies exponentially with the channel length. This can be seen from Fig. 6 where for different values of $\varepsilon_{ox}$ the slope of the curve ($\Lambda$) remains the same. We also see the DIBL reduces with increasing dielectric constant. However, even for a high dielectric constant ($\varepsilon_{ox} = 100$), the reduction in DIBL is not significant for all practical purposes.

## IV. Comparison with Experiments

We use the experimental data from ref. [19], [20] to compare our simulations. We perform the calculations mentioned in Section I to extract the scale length for the experimental device structure. The device has a 5 nm $HfO_2$ as back gate oxide and 5 nm $Al_2O_3$ as top gate oxide. In these simulations, we use short contacts to best represent electrostatics of the real device. The device schematic is shown in Fig. 7a. Using simulations, we plot in Fig. 7b the scale length for different $t_{ch}$. We can also define and extract a minimum channel length ($L_{min}$) below which the device DIBL is more than 100 mV/V. The $L_{min}$ is also plotted in Fig. 7b. For $t_{ch} \sim 0.6$ nm, $\Lambda \approx 6$ nm and the $L_{min} \approx 22$ nm. Note that $\Lambda$ is slightly different than our previous analytical expression of $4/3 t_{ox} = 7.5$ nm because the device geometry and contacts are different than the ideal symmetric DG FET design. As seen from Fig. 7b, we also note that $L \sim 3.5\Lambda$ instead of the usually accepted $3\Lambda$ [4].

We further compare the scale length for short and tall contacts in Fig. 7c. For the short contact, the impact of the drain field on the channel is smaller compared to the tall contact. As a result, the scale length with short contact is smaller. In Fig. 7d, we calculate DIBL as a function of channel length for short and tall contacts. The experimental results match quite well with the short contact as seen in Fig. 7d, which supports our theory.

We can use the theory developed in this paper to design a hypothetical 2D FET with 5 nm channel length and negligible SCE. The best electrostatic for such a transistor can be achieved by constructing a symmetric DG FET ($\gamma = 4/3$) which should have a $\Lambda = L/3.5 \sim 1.4$ nm or smaller. As per our simulations, this implies that the gate oxide thickness for the 2D FET should be smaller than 1.1 nm. As we wish to scale down the transistor to even smaller channel lengths, we also need to scale the oxide thickness, which will be eventually limited by the gate leakage current.

## V. Conclusion

We have established the scale length theory for 2D FETs. We show that for 2D DG FETs, $\Lambda$ scales linearly with $t_{ox}$ or $\Lambda \sim 4/3 t_{ox}$. The oxide dielectric constant does not affect the scaling in DG FETs. Most importantly, the well know scaling length expression $\Lambda \approx \sqrt{\epsilon_{ch} t_{ch} t_{ox}/\epsilon_{ox}}$ used to design modern transistors is incorrect for transistors with $t_{ch} <$ 4 nm. The oxide dielectric constant and the surrounding dielectric constant still play a minor role in scale lengths for SG FETs. Nonetheless, the scale length in SG FETs will always be larger than symmetric DG FETs for the same gate oxide thickness. We have also uncovered that in addition to an exponential dependence on the scale length, the DIBL for 2D FETs has a sub-linear dependence on the oxide dielectric constant.

Eventually, we use the scaling theory to explain a surprisingly high DIBL of 120 mV/V in a 20 nm channel length 2D DG FET. We have also found that the device electrostatic can be affected by the contact geometry. To avoid SCEs, the 2D transistors below 5 nm channel lengths will need a gate oxide thinner than 1.1 nm.

## Acknowledgments

We acknowledge technical discussion with Dr. Paul Solomon from IBM (Yorktown Heights). Authors would also like to acknowledge Isha Datye for help with proofreading the manuscript. This work has been partly supported by the NCN-NEEDS program, which is funded by the NSF contract 1227020-EEC and by the Semiconductor Research Corporation (SRC). The study was partly supported by the NSF EFRI 2-DARE grant 1542883, by the AFOSR grant FA9550-14-1-0251, and by the Systems on Nanoscale Information fabriCs (SONIC, one of six SRC STARnet Centers sponsored by MARCO and DARPA).

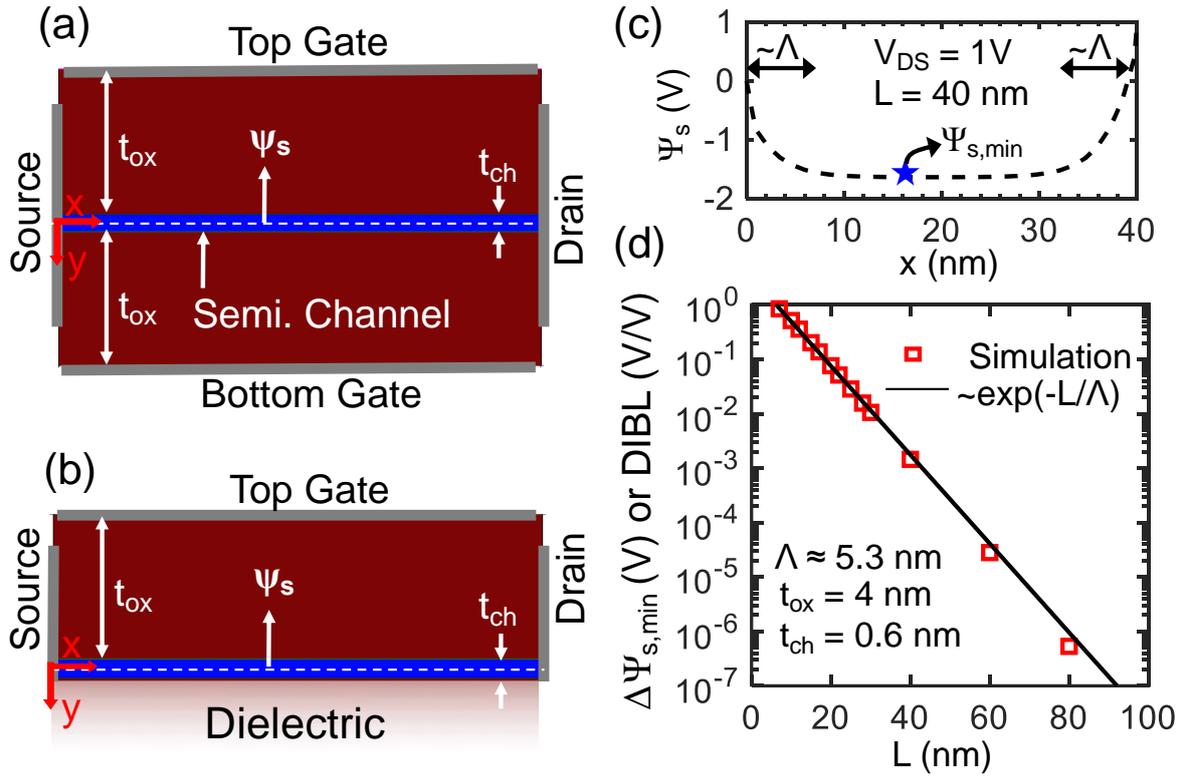

**Figure 1**: **Methodology**. (a) DG FET structure and (b) SG FET structure used to extract scale length ($\Lambda$) using TCAD simulations. For all simulations, $V_{GS}$ = -1 V and $V_{DS}$ = 1 V. For both structures, the electrostatic potential $\psi_s$ is calculated along the dashed white line which is along the path of maximum current flow. We perform finite element simulation on DG FET with $t_{ox}$ = 4 nm and $t_{ch}$= 0. 6 nm, which is equivalent to the thickness of the single layer MoS$_2$. (c) $\psi_S$ vs. $x$ for DG FET with $L$ = 40 nm and calculated $\psi_{s,min}$ shown by blue star. (d) The difference between the minimum potential at each channel length with the minimum potential at a channel length of 500 nm ($\Delta\psi_{s,min} = \psi_{s,min}(L = 500 \text{ nm}) - \psi_{s,min}(L)$) is shown in the figure using red square symbols. $\Delta\psi_{s,min}$ is also known as drain induced barrier lowering or DIBL. The black bold lines show the exponential fit to the symbols.

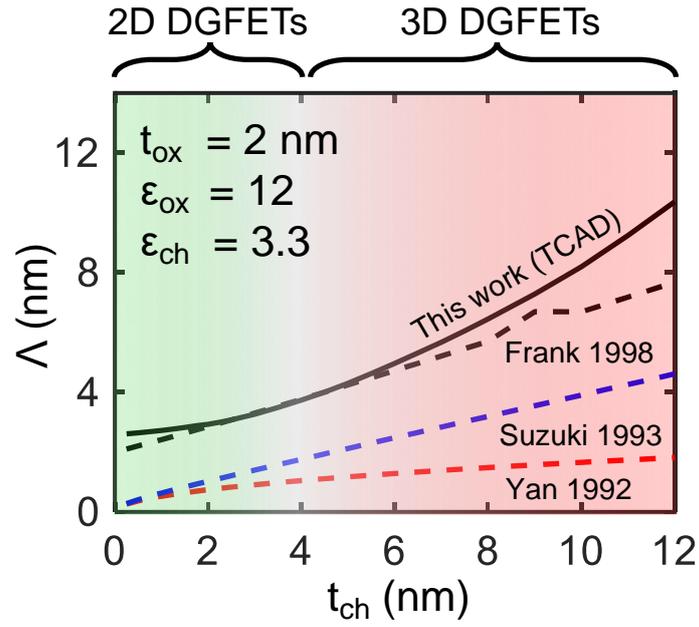

**Figure 2:** Comparison of scale lengths ($\Lambda$) calculated by Frank *et al.* [7], Suzuki *et al.* [6], and Yan *et al.* [8] For the sake of consistency, we keep the dielectric constant of the semiconductor the same with thickness.

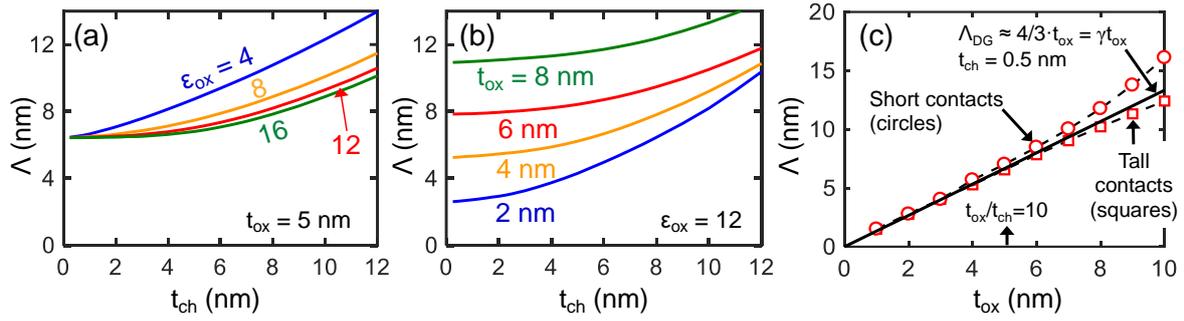

**Figure 3: Symmetric double-gate (DG) FET**: (a) Extracted scale length for DG FET versus the semiconductor channel thickness. The calculations are done for different oxide dielectric constants. (b) Extracted scale length for DG FET versus the semiconductor channel thickness for different oxide thickness. (c) Scale length vs. the physics gate-oxide thickness for short and long contacts. For the two extreme cases of the contact architecture, the scale length shows a linear trend with gate-oxide thickness with a slope ($\gamma$) of 4/3.

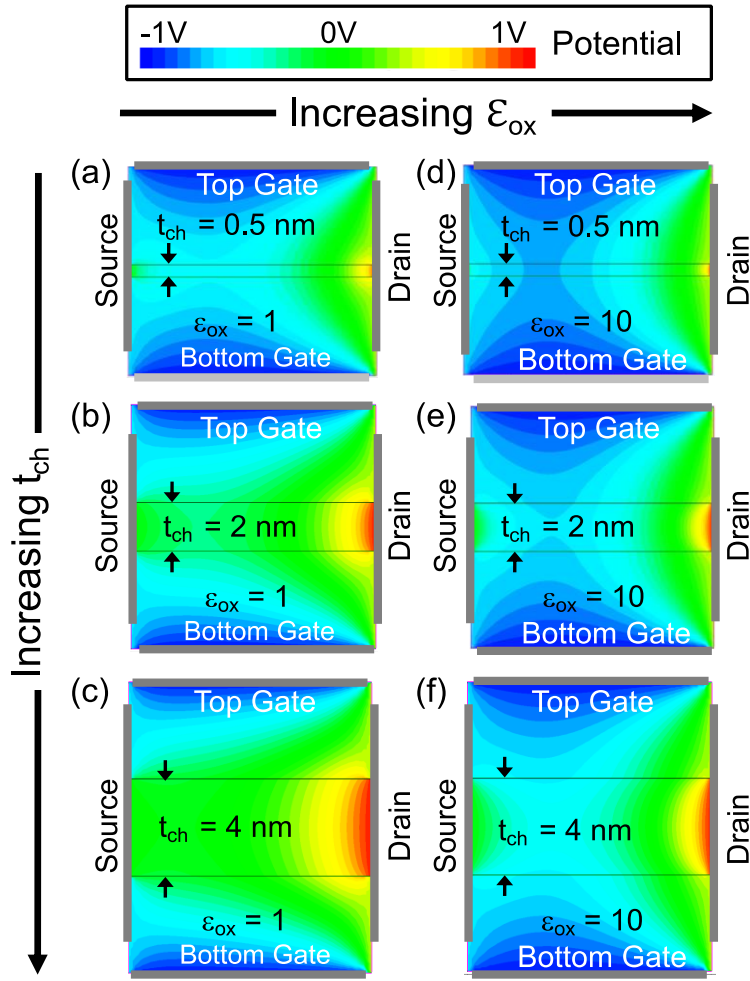

**Figure 4: Electrostatic potential in DG FET for different $t_{ch}$ and $\varepsilon_{ox}$.** (a) to (c) $t_{ch}$ increases from 0.5 nm to 4 nm, with $\varepsilon_{ox}$ = 1. (d) to (f) $t_{ch}$ increases from 0.5 nm to 4 nm, with $\varepsilon_{ox}$ = 10. For all the structures, $V_{GS}$ = -1 V and $V_{DS}$ = 1 V. The legend at the top shows the potential.

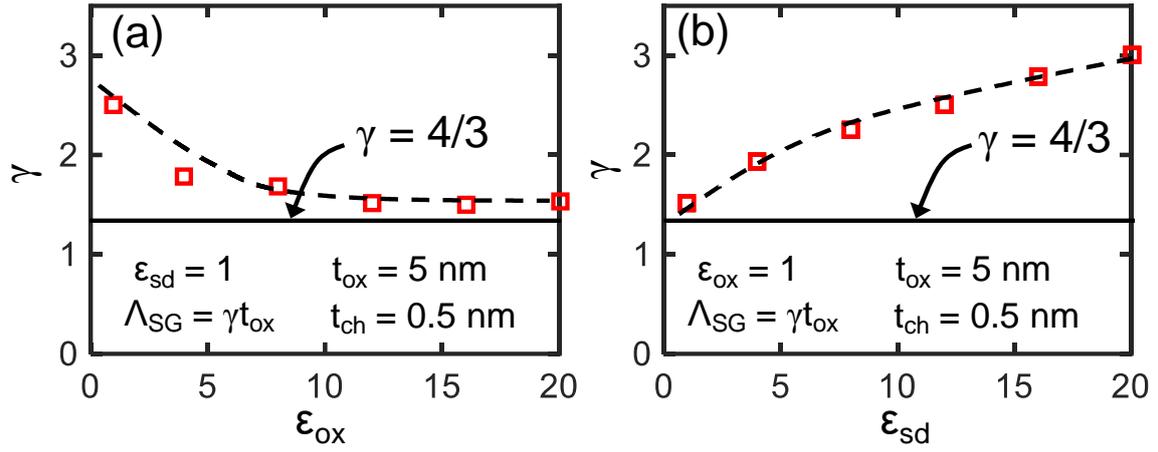

**Figure 5: Single-gate (SG) FET:** (a) $\gamma$ versus oxide dielectric constant ($\varepsilon_{ox}$) keeping the surrounding medium dielectric constant ($\varepsilon_{sd}$) fixed. (b) $\gamma$ versus surrounding dielectric constant ($\varepsilon_{sd}$) keeping the oxide dielectric constant ($\varepsilon_{ox}$) fixed. The dotted lines are drawn as guides for the eye.

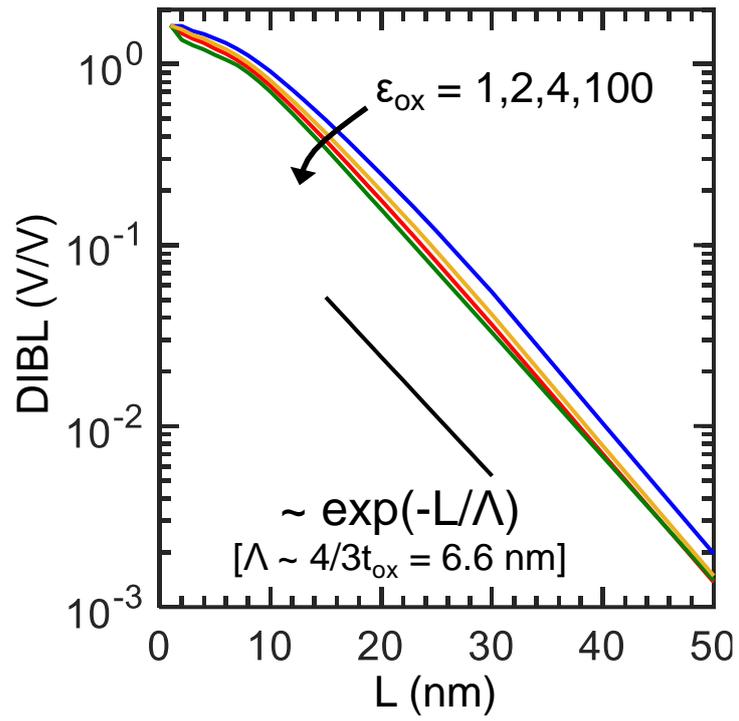

**Figure 6: Extracted DIBL for a symmetric double-gate FET** for various values of oxide dielectric constant with $t_{ox}$ = 5 nm and $t_{ch}$ = 0.6 nm.

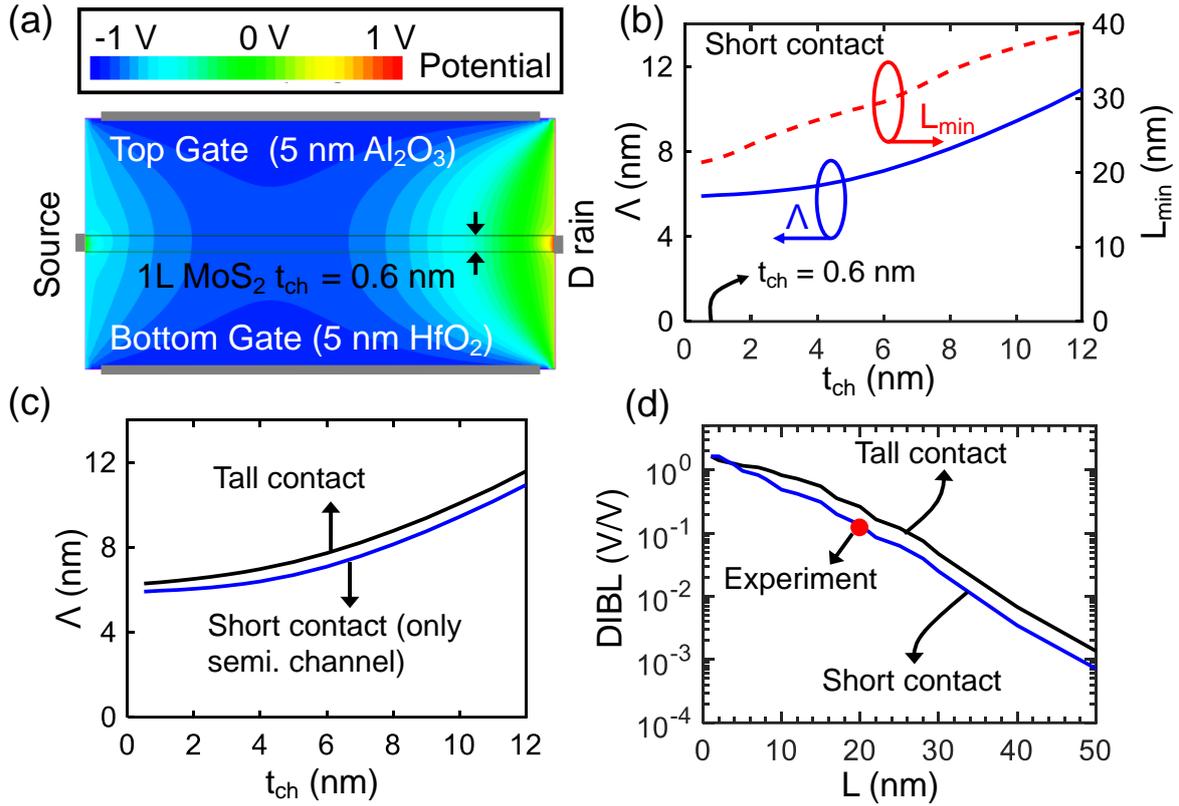

**Figure 7: Comparison between theory and experimental data.** (a) Electrostatics in a structure representing the experimental device from ref [19]. (b) The plot shows extracted $L_{min}$ and $\Lambda$ for the structure in (a) for varying channel thickness ($t_{ch}$). (c) Comparison of scale length extracted for short and tall contacts. (d) Extracted DIBL for the experimental device as a function of the channel length for short and tall contacts.